\documentclass[aps,pra,superscriptaddress,reprint,longbibliography]{revtex4-2}
\usepackage{CJK}
\usepackage{braket}
\usepackage{amsmath}
\usepackage{amssymb}
\usepackage{graphicx}
\usepackage{float}
\usepackage{hyperref}
\usepackage{xcolor}
\usepackage{verbatim}

\usepackage{multirow}
\newcommand{\R}{\tilde{\rho}}
\newcommand{\Rss}{\tilde{\rho}^{ss}}

\begin{document}
\begin{CJK*}{GB}{}
\newcommand{\floor}[1]{\lfloor #1 \rfloor}
\title{Synchronization Lower Bounds the Efficiency of Near-Degenerate Thermal Machines}
\author{Taufiq Murtadho}
\affiliation{Center for Theoretical Physics of Complex Systems, Institute for Basic Science (IBS), Daejeon 34126, Republic of Korea.}
\affiliation{Basic Science Program, Korea University of Science and Technology, Daejeon 34113, Republic of Korea.}
\affiliation{School of Physical and Mathematical Science, Nanyang Technological University, Singapore 639798, Singapore}
\author{Juzar Thingna}
\email[]{juzar$_$thingna@uml.edu}
\affiliation{Center for Theoretical Physics of Complex Systems, Institute for Basic Science (IBS), Daejeon 34126, Republic of Korea.}
\affiliation{Basic Science Program, Korea University of Science and Technology, Daejeon 34113, Republic of Korea.}
\affiliation{Department of Physics and Applied Physics, University of Massachusetts, Lowell, MA 01854, USA.}
\author{Sai Vinjanampathy}
\email[]{sai@phy.iitb.ac.in}
\affiliation{Department of Physics, Indian Institute of Technology-Bombay, Powai, Mumbai 400076, India.}
\affiliation{Centre of Excellence in Quantum Information, Computation, Science and Technology, Indian Institute of Technology Bombay, Powai, Mumbai 400076, India.}
\affiliation{Centre for Quantum Technologies, National University of Singapore, 3 Science Drive 2, Singapore 117543, Singapore.}

\date{\today}
\begin{abstract}
We study the relationship between quantum synchronization and the thermodynamic performance of a four-level near-degenerate extension of the Scovil--Schulz-DuBois thermal maser. We show how the existence of interacting coherences can potentially modify the relationship between synchronization and the coherent power output of such a maser. In particular, the cooperation and competition between interacting coherences, causes the coherent heat and efficiency to be bounded by the synchronization measure in addition to the well-studied power synchronization bound. Overall, our results highlight the role of quantum synchronization in the working of a thermal machine.
\end{abstract}
\maketitle
\end{CJK*}
\section{Introduction}
Quantum coherence has been a vital resource~\cite{StreltsovRMP17} in quantum thermodynamics and has been a topic of intense research in the recent past \cite{thermo_felix_book,goold2016role,millen2016perspective,sai2016quantum}. The presence of coherence typically boosts the performance metrics of thermal machines such as engines~\cite{SonPRXQ21}, refrigerators \cite{kosloff2014quantum}, and batteries \cite{campaioli2017enhancing,rossini2020quantum,julia2020bounds,SonPRA22}. Moreover, coherences, which can be thought of as phases, have also been subject to resource counting studies from an information theoretic perspective \cite{coh_noneq_freeen}. In the case of non-degenerate systems, that possess diagonal steady states in the absence of driving, such coherences are usually generated using a coherent drive. Such systems do not allow coherences to interact and thus a thorough understanding of the performance of quantum thermal machines with interacting coherences is lacking.

Classically, interacting phases are well studied in coupled Kuramoto models where phase pulling can produce a well-known second-order phase transition involving \textit{cooperative} phase locking called synchronization \cite{pikovsky2002synchronization}. Besides this, differences in coupling of the Kuramoto model can lead to \textit{competition} as well, giving rise to a variety of behavior such as anti-synchronization and chimera, a phenomenon explored even in the quantum regime~\cite{bastidas2015quantum}. Specifically, in quantum systems, such coupled phase oscillator models arise due to a variety of different reasons, one of them being degeneracies. Recently, for non-degenerate quantum systems coupled to diagonal baths, it was highlighted that synchronization measures fully account for all coherences in the system \cite{jaseem2020generalized} and can help explain the performance of quantum thermal machines. 

In contrast to coherently driven quantum systems, incoherently driven near-degenerate levels can generate bath-induced coherences. Since the energetic cost of generating coherences is (nearly) zero for making transitions between (near-)degenerate levels, such systems cause the thermodynamic analysis to be decoupled from synchronization. Moreover, the (near-)degeneracies cause such coherences to interact with each other and induce cooperation or competition between the different phases. 

In an accompanying manuscript, we show that synchronization of driven, dissipative \emph{exactly}-degenerate open quantum systems exhibit a crossover from cooperation to competition. In this manuscript, we highlight that this synchronous behavior aids in the understanding of nanoscale thermal machines. We show that while coherent driving of near-degenerate four-level thermal masers has a tendency to entrain the quantum system to the external drive, coupled coherences exert a simultaneous force that can be either competitive or cooperative. This interplay between cooperation and competition leads to a rich dynamical regime whose implications on the thermal maser are observed in terms of its performance metrics (power, heat, and efficiency) being bounded by the synchronization measure.

In Sec.~\ref{sec:fourlevel} below, we begin with a discussion of the near-degenerate four-level thermal maser. Then, we present an analysis of competition and cooperation in four-level thermal masers taking into account near-degeneracy and bath-induced coherence in Sec. \ref{sec:coexistence}. Then, in Sec. \ref{sec:thermObs} and Sec. \ref{sec:SyncThermoConnect} we connect the competition and cooperation to relevant thermodynamic quantities of the thermal machine, specifically its power and heat current. In \cite{jaseem2020quantum}, the coherent power output of a non-degenerate three-level maser was shown to be related to the measures of synchronisation and it was noted in \cite{solanki2022role} that this relationship does not hold for degenerate systems where there is no energetic cost of generating coherences. Furthermore such degeneracies were related to synchronisation blockade recently \cite{solanki2022symmetries}. We investigate this relationship in detail and connect synchronization to steady state heat current. The latter is then used to derive lower-bound on the efficiency of near-degenerate thermal machines in the mutual coupling dominant regime. Finally, we summarize our main results in Sec.~\ref{sec:summary}.
\section{Four-level Thermal Maser: Near degeneracy and noise-induced coherence}\label{sec:fourlevel}
\begin{figure}
    \centering
    \includegraphics[width =0.45\textwidth]{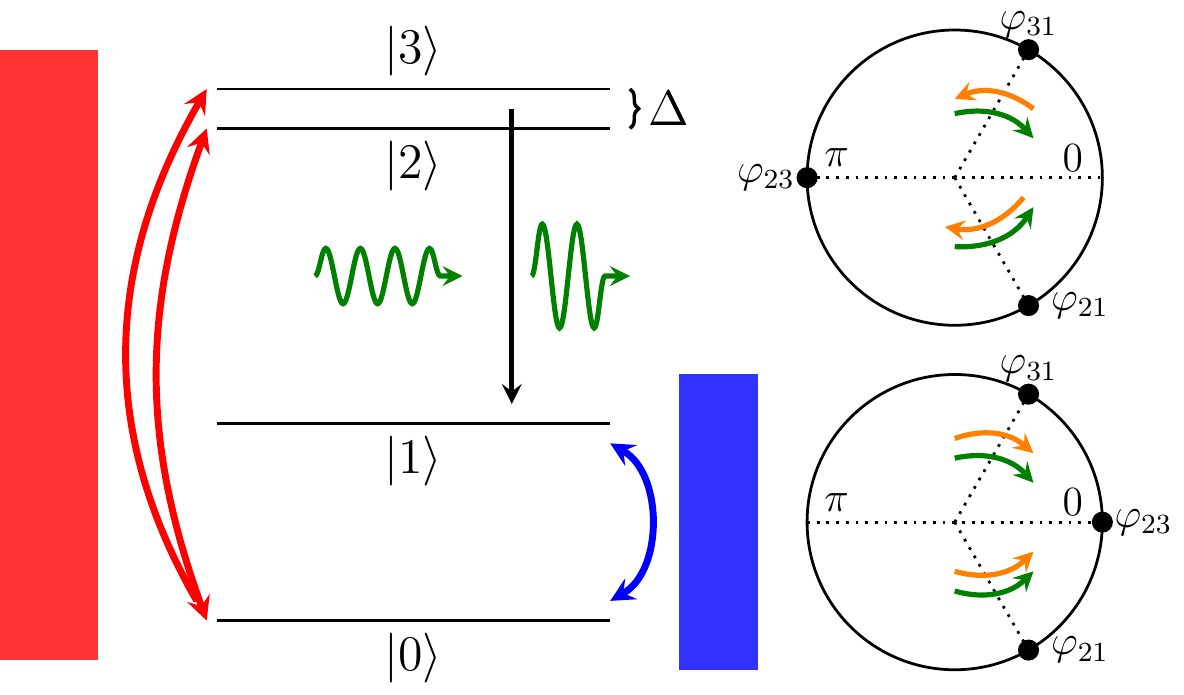}
    \caption{Schematic of the four-level Scovil--Schulz-DuBois thermal maser. Here,  $\Delta \equiv \omega_3 - \omega_2$ is the near-degenerate energy gap and $p$ is the noise-induced coherence strength that arises due to the hot heat bath causing interference between the near-degenerate levels and the ground state (red arrows). The relative phases $\varphi_{ij}$ between states $|i\rangle$ and $|j\rangle$ are depicted on the circles on the right and the arrows indicate the direction in which the non-degenerate phases ($\varphi_{31}$ and $\varphi_{21}$) move depending on the strength of mutual coupling (orange arrows) and entrainment (green arrows). In the engine (top circle) we observe both cooperation and competition, whereas in the refrigerator regime (bottom circle) we find only cooperating behavior.}
    \label{fig:four_level_maser}
\end{figure}
We begin with the analysis of a four-level thermal maser that operates under a temperature difference to create a population inversion as depicted in Fig.~\ref{fig:four_level_maser}. The original Scovil--Schulz-DuBois thermal maser is a minimal setup that comprises of three-levels and is one of the first models of a thermal heat engine~\cite{scovil1959three}, which has been recently extended to four levels~\cite{dorfman2018efficiency} and beyond~\cite{niedenzu2015performance, murtadho2023coexistence} to include effects of bath-induced coherence. As a generalization of this, we consider the four-level model consisting of a free Hamiltonian
\begin{equation}
H_0 = \omega_1\ket{1}\bra{1}+\sum_{j =2}^{3}\omega_j\ket{j}\bra{j},
\end{equation}
with $\omega_{3}>\omega_{2}>\omega_1>\omega_0 = 0$. The \emph{near-degenerate} levels ($\omega_2, \omega_{3}$) are spread over a small interval $\Delta \equiv \omega_{3} - \omega_{2} \ll (\omega_1-\omega_0)$ and $\Delta \ll (\omega_2 - \omega_1)$. The system is driven out-of-equilibrium by a hot (temperature $T_h$) and a cold bath (temperature $T_c$) as well as an external time-periodic drive of strength $\lambda$ and frequency $\Omega \approx (\omega_2 - \omega_1)$. The role of the baths is to establish population inversion between the near-degenerate manifold $\{\ket{2}, \ket{3}\}$ and the first-excited state $\ket{1}$. Specifically, when population inversion is achieved, the external drive can easily trigger stimulated emission, thereby producing power, i.e. the system acts as an engine. On the other hand, if there is no population inversion, the system absorbs power from the external drive, i.e. the system acts as a refrigerator.

When the baths are weakly coupled to the machine working fluid $H_0$, the reduced dynamics of the system is governed by a quantum master equation for the reduced density matrix $\rho$~\cite{dorfman2018efficiency} that reads,
\begin{equation}
\label{eq:QME}
\frac{d\rho}{dt} = -i[H_0+V(t), \rho]+\mathcal{D}_h[\rho]+\mathcal{D}_c[\rho].
\end{equation}
The operator $V(t)$ is a collective drive \cite{murtadho2023coexistence},
\begin{equation}
\label{eq:collective_drive}
V(t)  = \lambda e^{i\Omega t}\sum_{ j= 2}^{3}\ket{j}\bra{1}+\text{h.c.},
\end{equation}
that stimulates transitions between all the near-degenerate energy levels and the first-excited state. The cold-bath dissipator $\mathcal{D}_c[\rho]$ takes the Gorini-Kassakowski-Sudarshan-Lindblad (GKSL)~\cite{GKS76,Lindblad76} form
\begin{equation}
\label{eq:cold_dissipator}
\mathcal{D}_c[\rho] = \sum_{\mu = 1}^{2}\Gamma_{c_\mu}(2c_\mu\rho c_{\mu}^\dagger - \{c_{\mu}^\dagger c_\mu, \rho\}),
\end{equation}
with jump operators $c_1 = c_2^\dagger = \ket{0}\bra{1}$ and the decay rates satisfying local detailed balance $\Gamma_{c_1} = \gamma_c(1+n_c)$ and $\Gamma_{c_2} = \gamma_c n_c$. Here, $\gamma_c$ is the coupling strength squared between the system and the cold bath. 

The hot-bath dissipator $\mathcal{D}_h[\rho]$ connects the first excited state to the near-degenerate manifold. The near-degeneracy causes a breakdown of the secular approximation~\cite{breuer2002theory} and results in a Bloch-Redfield form of the dissipator~\cite{bloch1957generalized,redfield1957theory,agarwal2001quantum,tscherbul2015partial,dorfman2018efficiency},
\begin{equation}
\label{eq:hot_dissipator}
\mathcal{D}_h[\rho] = \sum_{\mu =1}^{2}\sum_{i,j = 2}^{3} \Gamma_\mu^{ij} [h_\mu^{i} , \rho \, h_\mu^{j\dagger}]+ \Gamma_\mu^{ji} [ h_{\mu}^{i}\, \rho,h_{\mu}^{j\dagger}].
\end{equation}
The operators $h_{1}^{i} = h_2^{i} = \ket{0}\bra{i}$ and pair-wise decay rates $\Gamma_{1}^{ij} = P_{ij}\sqrt{\gamma_h^i\gamma_h^j}\,(1+n_h^{(j)})$ and $\Gamma_{2}^{ij} = P_{ij}\sqrt{\gamma_h^i\gamma_h^j}\, n_h^{(j)}$. Here, $\gamma_h^{i}$ denotes the squared coupling strength of the bath that induces transitions between the ground state $\ket{0}$ and the $i$th state of the near-degenerate manifold $\ket{i}$ ($i=2,3$). The mean-bosonic hot-bath distribution $n_h^{(i)} = (\exp[\omega_i/T_x]-1)^{-1}$ encodes information about the hot bath temperature $T_h$ (similar definition for the cold-bath with $\omega_i = \omega_1$). In this paper, we will throughout set $\gamma_h^{(2)}  = \gamma_h^{(3)} = \gamma_h$ and work with the Redfield form that despite not being positive in all parameter ranges~\cite{JThingaPHaenggi2012,ThingnaPRE13} provides accurate results when used appropriately in the weak system-bath coupling regime~\cite{Strunz20, BeckerPRL22}. 

The coefficients $P_{ij}$ are elements of the symmetric \emph{correlation matrix}~\cite{marsaglia1984generating}
\begin{equation}
P_{ij} = \begin{cases}
1 & \text{if } i = j\\
p_{ij} &\text{for } i \neq j.
\end{cases}
\end{equation}
where $|p_{ij}|\leq 1$. Above since we are focused on a four-level thermal maser, the matrix $P$ is a $2 \times 2$ matrix with its off-diagonal elements $|p|\leq 1$. The elements $p$ originate from the dipole-alignment factor~\cite{agarwal2001quantum,tscherbul2015partial} with $p = \mathbf{d_2}\cdot \mathbf{d_3}/|\mathbf{d_2}||\mathbf{d}_3|$ such that $\mathbf{d}_i = \braket{i|\mathbf{d}|i}$ and $\mathbf{d}$ being the dipole operator. Intuitively, $|p|$ is the strength of quantum interference between thermalization pathways $\ket{0}\leftrightarrow \ket{2}$ and $\ket{0}\leftrightarrow \ket{3}$ while its sign encodes whether the interference is constructive or destructive. The parameter $p$ determines the strength of the noise-induced coherence with $p=0$ yielding no coherence due to the hot bath and $|p|=1$ yielding maximum coherence. The above Redfield hot-bath dissipator takes the completely positive Lindblad form~\footnote{By `Lindblad' we refer to the general Lindblad form $d\rho/dt = -i[H,\rho] + \sum_{i,j} K_{ij} (2A_i \rho A_j^{\dagger} - \{A_j^{\dagger}A_i,\rho\}$ where $A_i \equiv h_i$ are the jump operators and $K_{ij}$ is the positive Kassakowski matrix.} when $\Gamma_\mu^{ij} = \Gamma_{\mu}^{ji}~ \forall ~i,j = 2,3$, which can be satisfied in the exact degenerate limit ($\Delta = 0$ wherein $n_h^{(i)} = n_h$). In this limit, the Kossakowski matrix becomes proportional to the correlation matrix and thus positivity can be ensured if the correlation matrix is positive definite, i.e., for $-1\leq p \leq 1$. When the correlation matrix has a zero eigenvalue ($p=\pm 1$), the dissipator possesses a dark state~\cite{gelbwaser2015power,Thingna16} that can lead to multiple steady states~\cite{Thingna_2020,Thingna21}.

The above quantum master equation~\eqref{eq:QME} can be transformed to a rotating frame such that any operator $\tilde{O} = \exp[i\tilde{H} t] O \exp[-i\tilde{H} t]$ with
\begin{equation}
\tilde{H} = \frac{\Omega}{2}\sum_{j=2}^{3}\ket{j}\bra{j} -\frac{\Omega}{2}\ket{1}\bra{1}.
\end{equation}
The above transformation eliminates the explicit time dependency from the coherent evolution and leaves the dissipators invariant leading to a quantum master equation in the rotating frame,
\begin{equation}
\label{eq:corot_QME}
\frac{d\tilde{\rho}}{dt} = -i[H_0-\tilde{H}+\tilde{V}, \tilde{\rho}]+\mathcal{D}_h[\tilde{\rho}]+\mathcal{D}_c[\tilde{\rho}],
\end{equation}
with $\tilde{V} = \lambda\sum_{j=2}^{3}\ket{j}\bra{1} + \text{h.c.}$. Throughout this work, we will focus our attention on the four-level thermal maser investigating the effects of near degeneracy $\Delta$ and the dipole-alignment factor $p$ (noise-induced coherence strength) on quantum synchronization and the thermodynamic observables of the thermal maser. In particular, unlike Ref.~[\onlinecite{murtadho2023coexistence}] we will not explore the effect of several degenerate levels (generalized Scovil--Schulz-DuBois thermal maser) and restrict ourselves to the easily tractable and physically intuitive four-level heat machine.

\section{Coexistence of entrainment and mutual coupling in a near-degenerate thermal maser}\label{sec:coexistence}
\begin{figure}
    \centering
    \includegraphics[width = 0.45\textwidth]{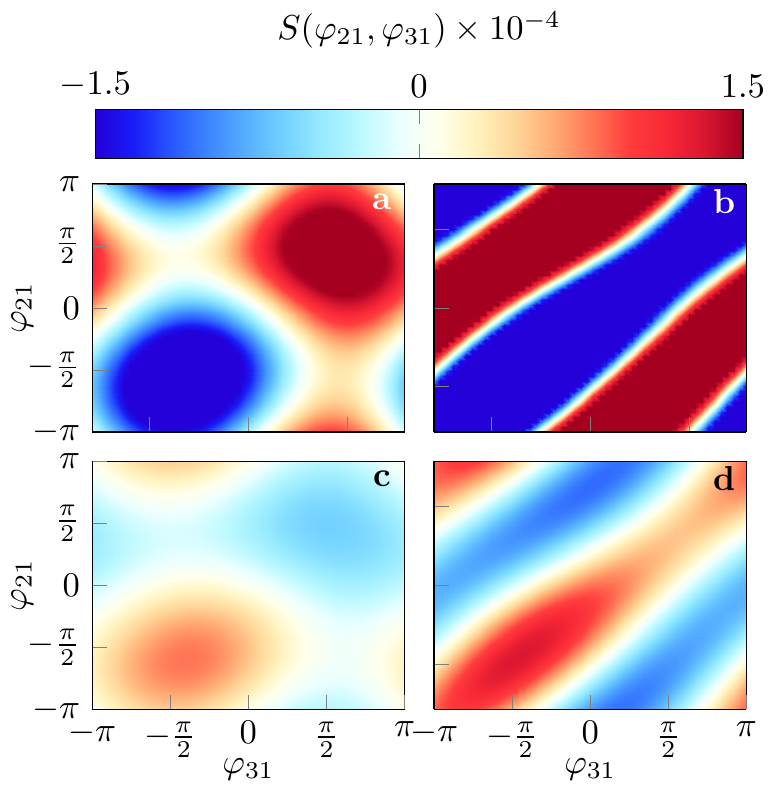}
    \caption{Quasi-probability phase distribution $S(\varphi_{31}, \varphi_{21})$ in the engine ({\bf a}-{\bf b}, $n_h^{(2)}/n_c = 5$) and refrigerator ({\bf c}-{\bf d}, $n_h^{(2)}/n_c = 0.5$) regime. The noise-induced coherence strength $p = 0.5$ in {\bf a} ($k = 4.5$) and {\bf c} ($k = 3.15$) and $p = -0.99$ in {\bf b} ($k = 0.03$) and {\bf d} ($k = 0.021$). Above, $\varphi_{ij} \equiv \phi_i - \phi_j$ is the phase difference associated with states $\ket{i}$ and $\ket{j}$. In {\bf a} and {\bf c}, both $\varphi_{21}$ and $\varphi_{31}$ are localized in the rotating frame, indicating entrainment to the external drive. Meanwhile, in {\bf b} and {\bf d}, the phases localize to a ring (on a torus), indicating that only the relative phase $\varphi_{23} = \varphi_{21} -\varphi_{31}$ is fixed. In this regime, synchronization is dominated by mutual coupling. We find that in the engine (refrigerator) regime, the mutual coupling synchronization is out-of-phase $\varphi_{23} \approx \pm \pi$ (in-phase $\varphi_{23} \approx 0$). Notice that there is a mirror symmetry between the phase distribution in the engine and refrigerator regime, i.e., the maximum (minimum) of $S({\varphi_{21}, \varphi_{31}})$ in the engine case is a minimum (maximum) in the refrigerator case. 
    The other model parameters are $\omega_2 = 3\omega_1$, $\Omega = \omega_2 - \omega_1+\Delta/2$, $\Delta = 0.05\omega_1$, $\gamma_h = \gamma_c = 0.1\omega_1$, $\lambda = 0.05\omega_1$, and $n_c = 0.1$.}
    \label{fig:phase_dist}
\end{figure}

The four-level thermal maser forms the minimal model in which entrainment and mutual coupling coexist. In the accompanying work~\cite{murtadho2023coexistence}, we show how competition and cooperation manifest in the synchronization measure for degenerate multilevel thermal masers. Depending on the thermodynamic functionality of the maser, the phases compete when the system behaves as an engine and cooperate when it acts as a refrigerator. 

The aim of this section is to show that the same phenomena persist even in presence of noise-induced coherence ($p \neq 0$) and near-degeneracy ($\Delta \neq 0$) by analytically examining the minimal model. We derive the formula for phase-space synchronization measure \cite{walter2014quantum} $S_{max}$ for the \emph{exactly} degenerate ($\Delta = 0$) four-level maser in presence of noise-induced coherence. This measure not only captures the strength of steady state coherences, but also its phase-matching condition \cite{murtadho2023coexistence}. The general $SU(D)$ quantum synchronization measure applicable to $D$-level systems is presented in Ref.~\cite{murtadho2023coexistence}. Here, we apply the general result for our specific case of $D = 4$, to obtain,
\begin{equation}
\label{eq:smax}
S_{max} = \frac{1}{16\pi^2}\times\begin{cases}
|\tilde{\rho}_{12}^{ss}|+|\tilde{\rho}_{13}^{ss}|+|\tilde{\rho}_{23}^{ss}| & n_c \leq n_h \\ 
|\tilde{\rho}_{12}^{ss}+|\tilde{\rho}_{13}^{ss}|-|\tilde{\rho}_{23}^{ss}| & n_h>n_c \; \& \; k>2 \\
\left(1+\dfrac{k^2}{2}\right)|\tilde{\rho}_{23}^{ss}| & n_h>n_c \; \& \; k\leq 2,\\
\end{cases}
\end{equation}
where $k = \gamma_h(1+n_h)(1+p)/\lambda$ is the \textit{dissipation-to-driving ratio}. The superscript of  $\tilde{\rho}$ denotes the steady state, whose analytical formula can be derived for $\Delta =0$ and $p \neq 0$ (see \hyperref[appendixA]{Appendix A}). The analytical solution for $\Rss$ can then be used to derive Eq. \eqref{eq:smax}  by following the recipe in the Supplementary Material of \cite{murtadho2023coexistence}. In fact, Eq. \eqref{eq:smax} has the same form as the result obtained in \cite{murtadho2023coexistence} with the exception that the dissipation-to-driving ratio $k$ now depends on the noise-induced coherence strength $p$. 

Equation~\eqref{eq:smax} displays cooperation and competition between entrainment and mutual coupling in different thermodynamic regimes.  The mutual-coupling (entrainment) contribution is represented by (non-) degenerate coherences $|\tilde{\rho}_{ij}^{ss}| (|\tilde{\rho}_{1j}^{ss}|)$ for $i,j=2,3$. In the refrigerator regime ($n_c<n_h$), entrainment and mutual coupling cooperate to increase the overall synchronization of the maser since all steady state coherences contribute positively to $S_{max}$. Meanwhile, in the engine case, we observe competition between coherences for $k<2$. The competition and cooperation is due to different phase configurations preferred by the two mechanisms, i.e., both prefer in-phase synchronization in the refrigerator case while one prefers in-phase and the other prefers out-of-phase in the engine case (see Fig.\ref{fig:phase_dist}). Moreover, in the engine case for $k<2$, synchronization is dominated by mutual coupling contribution $|\tilde{\rho}_{23}^{ss}|>|\tilde{\rho}_{12}^{ss}|+|\tilde{\rho}_{13}^{ss}|$. 

The $p$-dependence of $k$ allows us to explore different synchronization regimes by tuning the strength of noise-induced coherence. For example, in absence of noise-induced coherence ($p = 0$), the deep mutual coupling dominant regime $k\ll 2$ can only be explored in the strong-driving regime $\lambda \gg 2\gamma_h(1+n_h)$. However, if the driving is strong, one starts to deform the limit cycle and deviate away from the typical synchronization paradigm \cite{pikovsky2002synchronization}. Yet, in presence of noise-induced coherence ($p\neq 0$), the deep mutual coupling dominant regime of $k\ll 2$ can be easily explored in presence of total destructive interference ($p \rightarrow -1$). 

Figures \hyperref[fig:phase_dist]{2{\bf a}-{\bf b}} show the quasi-probability phase distribution $S(\varphi_{21}, \varphi_{31})$, where $\varphi_{ij} \equiv \phi_i - \phi_j$ is the relative phase between states $\ket{i}$ and $\ket{j}$. They are computed in the engine regime for $p = 0.5$ and $p = -0.99$. Figure \hyperref[fig:phase_dist]{2{\bf a}} shows the regime where entrainment is stronger than mutual coupling so that despite their competition, the phases localize in the rotating frame. On the other hand, Fig. \hyperref[fig:phase_dist]{2{\bf b}} shows the deep mutual-coupling dominant regime where entrainment is effectively lost but the relative phase is still fixed. Note that in both figures, the near-degenerate gap is set to be non-zero ($\Delta \neq 0$). 

\section{Thermodynamic Observables}\label{sec:thermObs}
In this section, we will derive the key thermodynamic observables, specifically steady state power and heat current following the standard recipe for weak-coupling thermodynamics~\cite{Alicki_1979}. We follow the standard energy partitioning~\cite{boukobza2006thermodynamics} that defines the internal energy as
\begin{equation}
E = \mathrm{Tr}(\tilde{\rho} H_0).
\end{equation}
The internal energy here is defined as that of the bare Hamiltonian $H_0$ and differs from the standard definition that involves the full Hamiltonian~\cite{Alicki_1979}. Importantly, such a definition remains the same in all pictures (Schr\"{o}dinger, Heisenberg, and Interaction) and thereby avoids the inconsistencies when one moves from one picture to another. Using the quantum master equation~\eqref{eq:corot_QME}, the change in internal energy can be expressed as,
\begin{align}
\label{eq:first_law}
\frac{dE}{dt} =  -i \mathrm{Tr}([H_0, \tilde{V}]\tilde{\rho}) + \mathrm{Tr}(\mathcal{D}_h[\tilde{\rho}]H_0)+\mathrm{Tr}(\mathcal{D}_c[\tilde{\rho}]H_0).
\end{align}

The energy flux is separated into three terms. The first term in \eqref{eq:first_law} is power and the following terms are heat fluxes from the hot and cold bath respectively \cite{boukobza2006thermodynamics,boukobza2007three},
\begin{eqnarray}
\label{eq:power}
P &=& -i \mathrm{Tr}([H_0, \tilde{V}]\tilde{\rho}) \nonumber \\
&=& 2\lambda\sum_{j=2}^{3} (\omega_j - \omega_1)\text{Im}(\tilde{\rho}_{1j}), \\
\label{eq:hot_heat_current}
\dot{Q}_h &=& \mathrm{Tr}(\mathcal{D}_h[\tilde{\rho}]H_0) \nonumber \\
&=& 2\gamma_h \Bigl(\sum_{j=2}^{3}\omega_j\left[n_h^{(j)}\tilde{\rho}_{00}-(1+n_h^{(j)})\tilde{\rho}_{jj}\right] \Bigr.\\
&&\Bigl.-[(1+n_h^{(2)})\omega_3 +(1+n_h^{(3)})\omega_2]p\, \mathrm{Re}(\tilde{\rho}_{23})\Bigr),\nonumber\\
\label{eq:cold_heat_current}
\dot{Q}_c &=& \mathrm{Tr}(\mathcal{D}_c[\tilde{\rho}]H_0)\nonumber \\
&=& 2\omega_1\gamma_c\left[n_c\tilde{\rho}_{00}-(1+n_c)\tilde{\rho}_{11}\right].
\end{eqnarray}
In the \emph{engine} regime, heat flows from the hot bath to the cold bath and power is produced in the steady state, i.e. $\dot{Q}_h^{ss} > 0,\; \dot{Q}_c^{ss} < 0, P^{ss}<0$. Conversely, in the \emph{refrigerator} regime, heat flows from the cold bath to the hot bath and power is consumed, i.e. $\dot{Q}_h^{ss} < 0,\; \dot{Q}_c^{ss} >0, \; P^{ss}>0$.

The heat current to the cold bath $\dot{Q}_c$ depends solely on populations whereas the heat from the hot bath $\dot{Q}_h$ depends on populations and coherences. Intuitively, since the hot bath connects the ground state and the near-degenerate manifold, a finite dipole alignment factor $p$ leads to noise-induced coherence causing the hot heat current to be dependent on both populations and coherences. Thus, the incoherent and coherent contribution to the hot bath's heat current~\cite{latune2019apparent, latune2019energetic} can be expressed as,
\begin{equation}
    \label{eq:inc_heat}
    \dot{Q}_h^{inc} = 2\gamma_h \sum_{j=2}^{3}\omega_j\left[n_h^{(j)}\tilde{\rho}_{00}-(1+n_h^{(j)})\tilde{\rho}_{jj}\right]
\end{equation}
\begin{equation}
\label{eq:coh_heat}
\dot{Q}_h^{coh} = -2\gamma_h[(1+n_h^{(2)})\omega_3 +(1+n_h^{(3)})\omega_2]p\, \mathrm{Re}(\tilde{\rho}_{23}).
\end{equation}
Note that the coherent heat current is proportional to $p$ which we set to be zero in the accompanying work \cite{murtadho2023coexistence}.

The coherent heat either suppresses or enhances the natural inclination of heat flow. For example, in the exactly degenerate ($\Delta = 0$) engine regime, a net heat flow coming from the hot bath to the system creates a population inverted state such that $\dot{Q}_h^{inc,ss}> 0$. From Eq.~\eqref{eq:coh_heat}, we see that the sign of $\dot{Q}_h^{coh,ss}$ depends on the sign of $p$ [$\because \text{Re}(\Rss_{23}) < 0$, see Eq.~\eqref{eq:deg_coherences}]. If noise-induced coherence strength $p>0$ ($p<0$), then the heat flow is enhanced (suppressed). Similarly, in the exactly degenerate ($\Delta = 0$) refrigerator regime, we observe enhancement or suppression of heat dumped into the hot bath due to either constructive ($p > 0$) or destructive ($p < 0$) interference.

\section{Connection between thermodynamic observables and synchronization}\label{sec:SyncThermoConnect}
In general the synchronization measure ($S_{max}$) and thermodynamic observables ($P$, $\dot{Q}_c$, and $\dot{Q}_h$) are unrelated. However, in the case of a three-level thermal maser, the steady state power of the maser is bounded by the synchronization measure~\cite{jaseem2020quantum} (power-synchronization bound) connecting two seemingly distinct quantities. In this section, we build the general framework of connecting the mathematically abstract notion of synchronization to physical thermodynamic observables for a four-level thermal maser. Despite several efforts analyzing the thermodynamic properties of the four-level thermal maser~\cite{dorfman2018efficiency, gelbwaser2015power, niedenzu2015performance}, there have so far been no connections between the synchronizing ability of such a thermal maser and its thermodynamic observables.

\subsection{Power - synchronization bound}\label{subsec:ps_bound}
\begin{figure}
    \centering
    \includegraphics[width = 0.45\textwidth]{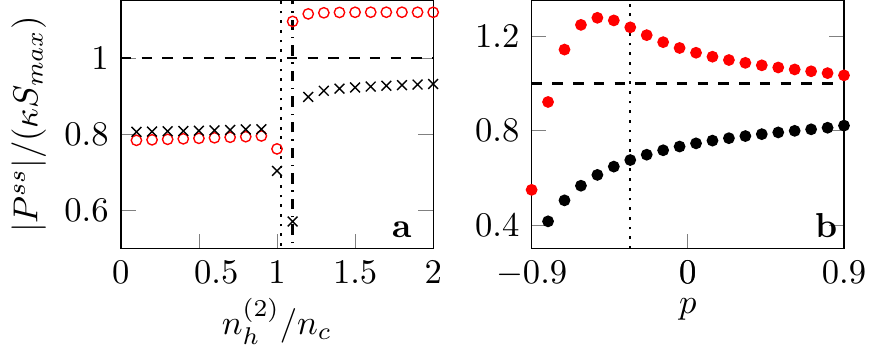}
    \caption{steady state power to the synchronization ratio $|P^{ss}|/(\kappa S_{max})$ with $\kappa = 32\pi^2\lambda\omega_{31}$ as a function of $n_h^{(2)}/n_c$ ({\bf a}) and noise-induced coherence parameter $p$ ({\bf b}). In {\bf a}, the near-degenerate energy gap $\Delta = 0.05\omega_1$ (red empty circles), $\Delta =  0.2\omega_1$ (black crosses), and the noise-induced coherence strength $p = 0.5$. In {\bf b}, the red filled circles are computed in the engine ($n_h^{(2)}/n_c = 5$) regime, and the black filled circles are computed in the refrigerator ($n_h^{(2)}/n_c = 0.5$) regime. The vertical dotted ($\Delta = 0.05\omega_1$) and dash-dotted ($\Delta = 0.2\omega_1$) lines in {\bf a} represent engine-to-refrigerator boundary defined by the change of sign in power. They slightly deviate from the degenerate case ($n_h^{(2)}/n_c = 1$). Meanwhile, the dotted line in {\bf b} marks the entrainment-dominant to the mutual-coupling dominant boundary ($k=2$). Power-synchronization bound is satisfied when $|P^{ss}|/(\kappa S_{max})\leq 1$ (below the dashed horizontal line) in both panels. Other parameter values are the same as Fig. \ref{fig:phase_dist}.}
    \label{fig:power_sync_bound}
\end{figure}
We begin by revisiting the relationship between quantum synchronization and the power of a near-degenerate thermal four-level maser. In the case of a three-level Scovil--Schulz-DuBois maser, the steady state power is bounded by the synchronization measure~\cite{jaseem2020quantum}, i.e.,
\begin{equation}
\label{eq:power_sync_bound}
|P^{ss}| \leq 2^{4}\pi\lambda(\omega_{2}-\omega_{1}) S_{max},
\end{equation}
also known as the power-synchronization (P-S) bound. The bound above has a different physical interpretation in the engine and refrigerator regimes: In the engine regime, P-S bound implies that the engine' power is limited by the amount the working substance entrains with the external drive. Whereas, in the refrigerator regime, since power is pumped into the system, the P-S bound suggests that there exists a maximum energy cost to ensure that the working substance is entrained to the external drive. In other words, in the engine regime, it can be inferred from the P-S bound that synchronization enhances the power output of the engine while in the refrigerator regime the P-S bound suggests that the external power supplied to the machine is utilized to synchronize the working substance.

We now investigate the P-S bound for a near-degenerate four-level thermal maser. Using Eq. \eqref{eq:power}, the power of the maser can be expressed as,
\begin{equation}
\label{eq:powerss}
    |P^{ss}| \leq 2\lambda\omega_{31}(|\tilde{\rho}_{12}^{ss}|+|\tilde{\rho}_{13}^{ss}|),
\end{equation}
where $\omega_{ij} \equiv \omega_{i} - \omega_{j}$. The steady state power $|P^{ss}|$ depends only on those coherences that are influenced by the external drive, whereas the synchronization measure $S_{max}$ is linearly dependent on all coherences. Specifically, the coherence $|\tilde{\rho}_{23}^{ss}|$, while it affects the synchronization measure $S_{max}$ [see Eq.~\eqref{eq:smax}], it does \textit{not} appear in the expression for steady state power [Eq.~\eqref{eq:powerss}]. Due to the additional dependence of $\Rss_{23}$, we find that the P-S bound is violated in the engine regime, while being respected in the refrigerator case, i.e., 
\begin{eqnarray}
\label{eq:ps_bound_ref}
|P^{ss} | &\leq 32\pi^2\lambda\omega_{31}S_{max} \quad \quad (n_c>n_h),\\
\label{eq:ps_bound_eng}
|P^{ss}|& \nleq 32\pi^2\lambda\omega_{31}S_{max} \quad \quad (n_h>n_c).
\end{eqnarray}
The above expression is analytically obtained in the exact degeneracy limit 
$\Delta = 0$. The violation originates due to the presence of the near-degenerate levels that induce competition between entrainment and mutual coupling \cite{murtadho2023coexistence}. The origin of the violation can be clearly seen from the expression for $S_{max}$ [Eq.~\eqref{eq:smax}] that in the engine regime for $k>2$ reads, $S_{max} \propto |\tilde{\rho}_{12}^{ss}|+|\tilde{\rho_{13}^{ss}}|-|\tilde{\rho}_{23}^{ss}|<|\tilde{\rho}_{12}^{ss}|+|\tilde{\rho_{23}^{ss}}|$. Although Eqs. \eqref{eq:ps_bound_ref} and~\eqref{eq:ps_bound_eng} are derived with the assumption $\Delta = 0$ (exact degeneracy), we show in Fig. ~\hyperref[fig:power_sync_bound]{\bf a-b} that the violation persists for small non-zero values of $\Delta$ where we plot steady state power to synchronization ratio $|P^{ss}|/(\kappa S_{max})$ with $\kappa = 32\pi^2\lambda\omega_{31}$ as a function of $n_h^{(2)}/n_c$ and noise-induced coherence strength $p$. The violation occurs if the ratio exceeds unity. 

In Fig. \ref{fig:power_sync_bound}\textbf{a}, we observe that other than a sharp discontinuity near the engine-to-refrigerator transition, $|P^{ss}|/(\kappa S_{max})$ does not strongly depend on $n_h^{(2)}/n_c$. We also find that the P-S bound is always satisfied in the refrigerator regime while being violated in the engine regime for small values of $\Delta$. Interestingly, as the near-degeneracy is further lifted by increasing $\Delta$, we find that the validity of the P-S bound is restored in both the engine and refrigerator regimes. 

We also plot $|P^{ss}|/(\kappa S_{max})$ as a function of noise-induced coherence strength $p$ in both the engine and refrigerator regime (see Fig. \ref{fig:power_sync_bound}\textbf{b}). We again observe that the bound is always satisfied in the refrigerator case while it is violated for most values of $p$ in the engine case. As $p \rightarrow -1$, the bound seems to be recovered again. However, this is not guaranteed in general. As $p\rightarrow - 1$ or equivalently $k\rightarrow 0$, the expression for $S_{max}$ only contains degenerate coherences while that of power only contains \textit{non}-degenerate coherences [see Eq. \eqref{eq:smax}]. Thus, as $p\rightarrow -1$, synchronization starts to decouple from power. 

The violation of the power-synchronization bound implies that in degenerate and near-degenerate four-level maser heat engines, more power can be generated than the upper bound set by synchronization. In contrast, in the refrigerator regime ($n_c>n_h$) the P-S bound is always satisfied but it is never saturated. This implies that synchronization can be generated with an energy cost less than the maximum cost imposed by the P-S bound.

\subsection{Coherent-heat -- Synchronization Bound}\label{subsec:qs_bound}
\begin{figure}
    \centering
    \includegraphics[width = 0.45\textwidth]{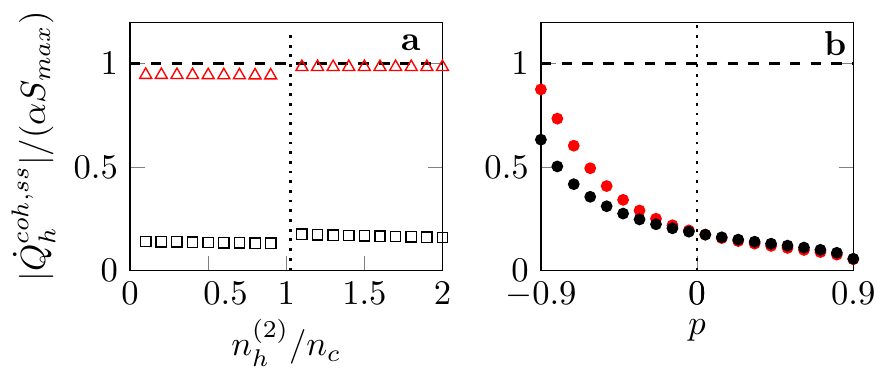}
    \caption{The coherent heat current to synchronization ratio $\dot{Q}_h^{coh,ss}/(\alpha S_{max})$ with $\alpha = (8\pi)^2\gamma_h\omega_3(1+n_h^{(2)})|p|$ as a function of $n_h^{(2)}/n_c$ ({\bf a}) and noise-induced coherence parameter $p$ ({\bf b}). The squares in {\bf a} are computed for $p = 0.5$ and the triangles are computed for $p = -0.99$. The red filled circles in {\bf b} are computed in the engine regime ($n_h^{(2)}/n_{c} = 5$) and the black filled circles are in the refrigerator regime ($n_h^{(2)}/n_c = 0.5$). All the data points lie below the dashed horizontal line which means Q-S bound is always satisfied. The dotted vertical line in {\bf a} represents the engine-to-refrigerator boundary and in {\bf b} it represents the transition from constructive (positive $p$) to destructive (negative $p$) interference. The near-degenerate gap $\Delta = 0.05\omega_1$. Other parameter values are the same as Fig.~\ref{fig:phase_dist}.}
    \label{fig:coh_heat_sync_bound}
\end{figure}
Intuitively, when the working substance of a machine synchronizes it achieves a low entropy state. It is therefore natural to expect that in order to maintain this state the entropy of the bath should increase. Thus, it is expected that synchronization will be coupled to heat. In this subsection, we will solidify this intuition and rigorously show the connections between heat and synchronization, leading to the heat-synchronization (Q-S) bound. 

Specifically, we have seen in Sec.~\ref{sec:thermObs} that in the presence of noise-induced coherence, the heat contribution from the hot bath can be separated into incoherent and coherent terms. We will now connect the coherent heat current to synchronization by noting the inequality
\begin{equation}
|\dot{Q}_h^{coh, ss}| \leq 4|p|\gamma_h(1+n_h^{(2)})\omega_3 \text{Re}(\tilde{\rho}_{23}^{ss}).
\end{equation}
The above inequality is derived from Eq.~\eqref{eq:coh_heat} by making use of the fact that $\omega_3\geq \omega_2$ and $n_h^{(2)}\geq n_h^{(3)}$. Note that in contrast to power which is independent of \textit{degenerate coherences} $|\tilde{\rho}_{23}^{ss}|$, the coherent heat current is \textit{only} a function of degenerate coherences. Recall, also that synchronization measure $S_{max}$ is a function of both degenerate and non-degenerate coherences. One may then relate coherent heat current in the steady state with the synchronization measure. In the limit of $\Delta = 0$, we can use Eq.~\eqref{eq:smax} to derive a coherent heat-synchronization (Q-S) bound,
\begin{equation}
|\dot{Q}_h^{coh,ss}| \leq (8\pi)^2|p|\gamma_h(1+n_h^{(2)})\omega_3 S_{max}.
\end{equation}
The validity of this bound is trivial in the refrigerator regime where $S_{max}$ is given by the $\ell_1$-norm $C_{\ell_1}$ and noting that $\text{Re}(\tilde{\rho}_{23}^{ss})\leq |\tilde{\rho}_{23}^{ss}|\leq C_{\ell_1}$. Even in the engine regime, when mutual coupling dominates ($k<2$), the bound is easily satisfied since $\text{Re}(\tilde{\rho}_{23}^{ss})\leq |\tilde{\rho}_{23}^{ss}|\leq (1+k^2/2)|\tilde{\rho}_{23}^{ss}|$. Yet, perhaps more surprisingly, this bound is also valid in the engine entrainment dominant case ($k>2$). In this regime, since $S_{max} \propto (2k-1)|\tilde{\rho}_{23}^{ss}|\geq 3|\tilde{\rho}_{23}^{ss}|$ and $\dot{Q}^{coh,ss}_h \propto \mathrm{Re}(\tilde{\rho}_{23}^{ss}) \leq |\tilde{\rho}_{23}^{ss}|$, the Q-S bound is always satisfied. In other words, unlike the P-S bound which can be violated in presence of near-degeneracy, the Q-S bound is \textit{always} satisfied for $\Delta = 0$. 

Although we derived the bound using the exact degeneracy assumption ($\Delta = 0$), we show numerically in Fig.~\ref{fig:coh_heat_sync_bound} that the bound is still valid for small values of $\Delta\neq 0$. In Fig.~\ref{fig:coh_heat_sync_bound}{\bf a}, we plot the coherent heat to synchronization ratio $|\dot{Q}_h^{coh,ss}|/(\alpha S_{max})$ with $\alpha = (8\pi)^2 \gamma_h\omega_3|p|(1+n_h^{(2)})$ for different values of $n_h^{(2)}/n_c$ and noise-induced coherence strength $p$. We observe that $|\dot{Q}_{h}^{coh,ss}|/(\alpha S_{max})$ does not depend strongly on $n_h^{(2)}/n_c$ except for a discontinuity near engine-to-refrigerator boundary. Furthermore, we find that the bound is always satisfied and is not strongly dependent on $\Delta$. However, the bound is tight only deep in the mutual coupling regime, e.g. when $p \rightarrow -1$. 


\subsection{Lower Bound on Efficiency and Coefficient of Performance (COP)}
Efficiency is one of the most well-known performance metrics for thermal engines. It is defined by $\eta = -P^{ss}/\dot{Q}_h^{ss}$. Previously, we have demonstrated that both steady state power $P^{ss}$ and the coherent part of heat current $\dot{Q}_h^{coh,ss}$ are connected to synchronization measure $S_{max}$. It is then natural to ask whether efficiency is also connected to synchronization. The efficiency can be computed analytically in the degenerate limit ($\Delta = 0$) and it is given by the standard expression $\eta = 1-(\omega_1/\omega_2)$ (see Appendix \ref{appendixB}), which is also obtained for the three-level thermal maser. Thus, the efficiency in the exact degenerate limit depends only on the system's energy scale and it is independent of the synchronization measure $S_{max}$ or any other parameters. However, this is no longer guaranteed in the near-degenerate case ($\Delta \neq 0$), where the steady state is not analytically solvable. In such cases, the efficiency is generally a function of all the parameters of the system, baths, and drive. 

It is well-known that the efficiency of a heat engine is upper-bounded by the Carnot efficiency. Here, we show that synchronization sets a \textit{lower} bound to the efficiency of a \textit{near-}degenerate thermal maser. We will use the violation of P-S bound and the validity of Q-S bound which have been demonstrated for the near-degenerate $\Delta \neq 0$ case in subsections \ref{subsec:ps_bound}-\ref{subsec:qs_bound}. Below, we state the efficiency-synchronization (E-S) bound as, 
\begin{equation}
\label{eq:eff_lower_bound}
\eta  \geq \frac{\kappa S_{max}}{\dot{Q}_h^{inc,ss}+\alpha S_{max}} . 
\end{equation}
The above inequality is derived under the assumption that P-S bound is violated $|P^{ss}|>\kappa S_{max}$ and the Q-S bound is respected $\dot{Q}_h^{inc,ss} +\dot{Q}_h^{coh,ss}\leq \dot{Q}_h^{inc,ss}+|\dot{Q}_h^{coh,ss}| \leq \dot{Q}_h^{inc,ss}+\alpha S_{max}$ with $\kappa = 32\pi^2\lambda\omega_{31}$ and $\alpha = (8\pi)^2 |p|\gamma_h(1+n_h^{(2)})\omega_3$. Note that the incoherent heat current $\dot{Q}_h^{inc,ss}$ is only a function of populations [see Eq. \eqref{eq:inc_heat}] and so it is unrelated to synchronization. 

\begin{figure}
    \centering
    \label{fig5}
    \includegraphics[width = 0.45\textwidth]{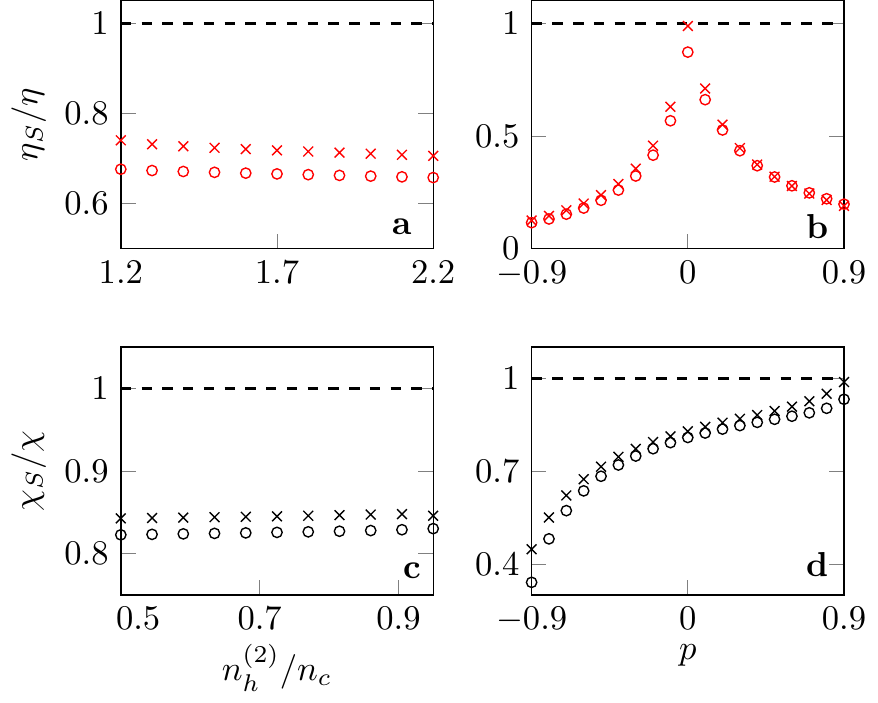}
    \caption{Synchronization lower bounds engine's efficiency $\eta$ (\textbf{a}-\textbf{b}) and refrigerator's COP $\chi$ (\textbf{c}-\textbf{d}) in near-degenerate four-level thermal maser. The red (black) data points are computed in the engine (refrigerator) regime. The circles (crosses) are computed with the value of near-degenerate energy gap $\Delta = 0.05\omega_1$ ($\Delta = 0.1\omega_1$). Panels \textbf{a}-\textbf{b} show efficiency ratio $\eta_S/\eta$ [see Eq.~\eqref{eq:effboundS}] as a function of $n_h^{(2)}/n_c$ (\textbf{a}) and noise-induced coherence parameter $p$ (\textbf{b}). Similarly, panels \textbf{c}-\textbf{d} show the COP ratio $\chi_S/\chi$ [see Eq.~\eqref{eq:COPboundS}] as a function of $n_h^{(2)}/n_c$ (\textbf{c}) and $p$ (\textbf{d}). In panels \textbf{a} and \textbf{c} the noise-induced coherence parameter is kept constant at $p = 0.1$. In panel \textbf{b} (\textbf{d}), the baths mean occupation number ratio is fixed $n_h^{(2)}/n_c = 2$ ($n_h^{(2)}/n_c = 0.5$). The bounds \eqref{eq:eff_lower_bound}-\eqref{eq:cop_smax_bound} are satisfied if all the data points lie below the dashed horizontal line. Other parameters are the same as in Fig. \ref{fig:phase_dist}}.  
    \label{fig:eff_cop_bound}
\end{figure}

We check the validity of the E-S bound Eq.~\eqref{eq:eff_lower_bound} for various parameter values in Fig. \hyperref[fig5]{5\textbf{a-b}}. Similar to the P-S and Q-S bound, we find that the ratio $\eta_S/\eta$ with 
\begin{equation}
\label{eq:effboundS}
 \eta_S = \frac{\kappa S_{max}}{\dot{Q}_h^{inc,ss}+\alpha S_{max}}, 
\end{equation}
does not strongly depend on the baths mean occupation number ratio $n_h^{(2)}/n_c$ (Fig. \hyperref[fig5]{5\textbf{a}}). Instead, it strongly depends on the noise-induced coherence parameter $p$ (Fig. \hyperref[fig5]{5\textbf{b}}). In the limit when there is no interference effect from the hot bath, i.e., $p \rightarrow 0$, we find that the bound is saturated.  

Similarly, the coefficient of performance (COP) in the \emph{exact} degenerate ($\Delta =0$) refrigerator regime can be analytically computed to be $\chi = \omega_1/(\omega_2 - \omega_1)$ (see Appendix \ref{appendixB}). In the near-degenerate ($\Delta \neq 0$) case, the COP is lower-bounded by the inverse of synchronization measure $S_{max}$,
\begin{equation}
    \label{eq:cop_smax_bound}
    \chi = \frac{\dot{Q}_c^{ss}}{P^{ss}} \geq \frac{\dot{Q}_c^{ss}}{\kappa S_{max}}.
\end{equation}
The lower bound is valid because the consumed power is always upper-bounded by $S_{max}$ in the refrigerator regime [see Eq. \eqref{eq:ps_bound_ref}]. We also note that $\dot{Q}_c^{ss}$ is only a function of populations [see Eq. \eqref{eq:cold_heat_current}] and hence is unrelated to the synchronization. 

We check the validity of the bound Eq.~\eqref{eq:cop_smax_bound} for various values of parameters in Fig. \hyperref[fig5]{5\textbf{c-d}}. Similar to the E-S bound, we find that the ratio $\chi_S/\chi$, with 
\begin{equation}
\label{eq:COPboundS}
    \chi_S = \frac{\dot{Q}_c^{ss}}{\kappa S_{max}},
\end{equation}
 does not strongly depend on the baths mean occupation number ratio $n_h^{(2)}/n_c$ (Fig. \hyperref[fig5]{5\textbf{c}}), but depends strongly on the noise-induced coherence parameter $p$ (Fig. \hyperref[fig5]{5\textbf{d}}). However, different from the E-S bound, the bound \eqref{eq:cop_smax_bound} is saturated in the limit $p \rightarrow 1$, i.e., when the hot bath causes constructive interference. 
\section{Summary and Discussion}\label{sec:summary}
There is intense debate regarding the optimal design of quantum thermal machines. Almost all design paradigms assume the systems are well modeled by Markovian master equations and furthermore pursue coherence from an information-theoretic resource cost point of view or a non-equilibrium cost point of view. While such analyses are definitely sufficient for non-degenerate systems, secular approximation begins to fail to describe near-degenerate systems interacting with a bath. Furthermore, interacting coherences can be well understood under the paradigm of coupled oscillator models, an analysis that has tremendously benefited classical dynamical systems. 

In this manuscript, we argue fundamentally that such a coupled oscillator model is indispensable to understanding the subtle dynamics of interacting coherences, which arise due to a variety of situations like those described in this manuscript. We developed the synchronization dynamics for a four-level atom coupled to two baths, generalizing the canonical thermal maser analysis by Scovil and Schulz-DuBois. We show that coherences interact with each other in two distinct qualitative ways. The first of these is a cooperative phase-locking phenomenon understood as entrainment dynamics of coherences being driven by an external field. Such entrainment dynamics and their contribution to the performance of a thermal maser were discussed earlier~\cite{jaseem2020quantum}. In this manuscript, we add to this entrainment dynamics and highlight the competition between coherences that can arise due to the bath coupling in our model. This coexistence of cooperation and competition gives rise to a richer tapestry of dynamics, uncoupling the power output from the thermal maser from the synchronization measure. The bound on power from the synchronization measure of the standard maser is modified to include coherent contributions to heat.

In presence of noise-induced coherence, in addition to the power-synchronization bound, we also find that coherent heat is bounded by synchronization measure showing that synchronization not only influences the useful work~\cite{HerpichPRX18, Ryu_2021} but can also bound the wasteful heat. The result of this work can be summarized in the following table namely,
\begin{center}
\begin{tabular}{ |c|c|c| } 
\hline
 &  Engine & Refrigerator \\
\hline
P-S bound & x &\checkmark \\
\hline
Q-S bound &\checkmark & \checkmark\\
\hline
E-S bound & \checkmark &\checkmark\\ 
\hline
\end{tabular}
\end{center}
where the check mark indicates that the bound is satisfied whereas the cross means the bound is violated. The lower bound of COP is included in the refrigerator column of E-S bound for compactness. Unlike the standard upper bounds like the Carnot efficiency, our analysis shows that efficiency can be \emph{lower} bounded in presence of quantum synchronization. In other words, a quantum synchronous working substance can boost the efficiency of the machine. Even though our analysis was limited to a four-level thermal maser, it would be an interesting future avenue to investigate other coupled oscillator models for interacting coherences to inform the future designs of quantum thermal machines.

\begin{acknowledgments}
This research was supported by the Institute for Basic Science in South Korea (IBS-R024-Y2). S.V. acknowledges support from a Government of India DST-QUEST grant number DST/ICPS/QuST/Theme-4/2019. The authors would like to thank V. Singh for the useful discussions.
\end{acknowledgments}

\bibliography{PRA_Murtadho.bib}

\appendix
\section{Steady state of Four-Level Thermal Maser with Interference}\label{appendixA}
The quantum master equation~\eqref{eq:corot_QME} can be expanded into a series of linear first-order differential equations for each density matrix element. We divide the elements into three groups: populations ($\rho_{ii}$ with $i=0,\cdots, 3$), non-degenerate coherences ($\rho_{12}$ and $\rho_{13}$), and degenerate coherences ($\rho_{23}$). In the following derivation, we have ignored the coherences $\rho_{0i}$ with $i = 1,2, 3$ since their dynamics are decoupled from other density matrix elements and they decay to zero in the steady state. 

We restrict ourselves to the exact degenerate case scenario with $\Delta=0$ and $n_h^{(i)} = n_h$ and obtain the equations for the populations, namely,
\begin{eqnarray}
    \label{eq:rho_11}
    \frac{d\R_{11}}{dt} &=& i\lambda\sum_{j=2}^{3}(\R_{1j} - \R_{j1})-2\xi_c \R_{11} +2\gamma_c n_c\R_{00}, \\
    \label{eq:rho_22}
    \frac{d\R_{22}}{dt} &=& -i\lambda(\R_{12}-\R_{21})-\xi_h p(\R_{23}+\R_{32})-2\xi_h \R_{22}\nonumber \\&&+2\gamma_hn_h\R_{00},\\
    \label{eq:rho_33}
    \frac{d\R_{33}}{dt} &=& -i\lambda(\R_{13}-\R_{31})-\xi_hp(\R_{23}+\R_{32}) -2\xi_h\R_{33}\nonumber \\
    &&+2\gamma_hn_h\R_{00}.
\end{eqnarray}
Above we have introduced $\xi_x = \gamma_x (1+n_x)$ with $x=c,h$ to simplify our notation. The equations for the non-degenerate and degenerate coherences read,
\begin{widetext}
\begin{eqnarray}
\label{eq:rho_12}
\frac{d\R_{12}}{dt} &=& i(\omega_2 - \omega_1 - \Omega)\R_{12}-(\xi_c+\xi_h)\R_{12} - \xi_h p\R_{13}+i\lambda\left(\R_{11} - \R_{22}-\R_{32}\right),\\
\label{eq:rho_13}
\frac{d\R_{13}}{dt} &=& i(\omega_2 - \omega_1 + \Delta -\Omega)\R_{13}-(\xi_c+\xi_h)\R_{13} -\xi_hp\R_{12}+i\lambda\left(\R_{11} - \R_{33} - \R_{23}\right),\\
    \label{eq:rho_23}
    \frac{d\R_{23}}{dt} &=& -i\Delta\R_{23}-2\xi_h\R_{23}-i\lambda(\R_{13} - \R_{21}) +2\gamma_h n_h p\rho_{00}-\xi_h p(\R_{22}+\R_{33}) .
\end{eqnarray}
\end{widetext}
To find an analytical steady state ($d\R_{ij}/dt = 0\; \forall i,j$) solution, we further restrict ourselves to the resonant drive $\Omega = \omega_2 - \omega_1$ and degenerate ($\Delta = 0$) limit. Using Eqs.~\eqref{eq:rho_12} and \eqref{eq:rho_13} and splitting them into their real and imaginary parts we obtain,
\begin{eqnarray}
    \label{eq:12+12*+13+13*}
    \text{Re}(\Rss_{12})+\text{Re}(\Rss_{13}) &=& 0,\\
    \label{eq:12+12*-13-13*}
    \left[\xi_c+\xi_h(1-p)\right]\text{Re}(\Rss_{12})+i\lambda\;\text{Im}(\Rss_{23}) &=& 0.
\end{eqnarray}
Furthermore, using Eq.~\eqref{eq:12+12*+13+13*} and computing the imaginary part of Eq.~\eqref{eq:rho_23} we obtain, 
\begin{equation}
    \label{eq:23-23*}
    2\xi_h\;\text{Im}(\Rss_{23}) + i\lambda \;\text{Re}(\Rss_{12}) = 0.
\end{equation}
Note that both Eqs.~\eqref{eq:12+12*-13-13*} and \eqref{eq:23-23*} are linear, involving the same density matrix elements. Therefore, combining these two equations yields, 
\begin{equation}
    \label{eq:simplify1}
\text{Re}(\Rss_{12}) = \text{Re}(\Rss_{13}) = \text{Im}(\Rss_{23}) = 0,
\end{equation}
i.e., the steady state (non-) degenerate coherences are real (imaginary). Further simplification can be sought by subtracting the imaginary part of Eq.~\eqref{eq:rho_12} from Eq.~\eqref{eq:rho_13} to yield,
\begin{equation}
    \left[\xi_c+\xi_h(1-p)\right] \left[\text{Im}(\Rss_{12})-\text{Im}(\Rss_{13})\right] +i\lambda(\Rss_{22} -\Rss_{33}) = 0.
\end{equation}
The equation above connects the difference between non-degenerate coherences and the population difference between the degenerate states. We can find another linearly independent equation by computing the population differences between degenerate states using Eqs.~\eqref{eq:rho_22} and \eqref{eq:rho_33}, i.e.,
\begin{equation}
i\lambda\left[\text{Im}(\Rss_{12})-\text{Im}(\Rss_{13})\right]+\xi_h(\Rss_{22}-\Rss_{33}) = 0.
\end{equation}
Combining the above two equations gives,
\begin{equation}
    \text{Im}(\Rss_{12}) = \text{Im}(\Rss_{13}) \qquad \rm{and}\qquad \Rss_{22} = \Rss_{33}.
\end{equation}
Thus, the population and the coherences associated with the degenerate states are equal. This makes intuitive sense because we are working in the exact degenerate limit. Using the constraints derived thus far together with a trace-preserving condition $\Rss_{00} = 1-\Rss_{11}-\Rss_{22}-\Rss_{33}$, the problem is reduced to four independent variables: $\Rss_{11}, \; \Rss_{22}, \; \Rss_{12},$ and $\Rss_{23}$. The corresponding four equations to solve for these variables can be derived from Eqs.~\eqref{eq:rho_11} -- \eqref{eq:rho_23} and they can be expressed as,
\begin{eqnarray}
    \label{eq:reduced1}
    2i\lambda \Rss_{12} - \xi_c\Rss_{11} + \gamma_c n_c(1-\Rss_{11}-2\Rss_{22}) &=& 0,\nonumber \\
    \label{eq:reduced3}
    \left[\xi_c+\xi_h(1+p)\right]\Rss_{12}-i\lambda\left(\Rss_{11}-\Rss_{22}-\Rss_{23}\right) &=& 0,\nonumber \\
    \label{eq:reduced2}
    i\lambda \;\Rss_{12} +\xi_hp\Rss_{23}+\xi_h\Rss_{22} - \gamma_h n_h(1-\Rss_{11}-2\Rss_{22}) &=& 0,\nonumber \\
    \label{eq:reduced4}
    i\lambda \Rss_{12} + \xi_h\Rss_{23} + \xi_h p\Rss_{22} -\gamma_h p n_h(1-\Rss_{11}-2\Rss_{22}) &=& 0.\nonumber\\
\end{eqnarray}
Solving the above linear system of equations for the four-independent steady state density matrix elements we first obtain the ratio between non-degenerate and degenerate coherence,
\begin{equation}
    \label{eq:coherence_ratio}
    \frac{\Rss_{12}}{\Rss_{23}} = i\frac{\gamma_h(1+n_h)(1+p)}{\lambda}. 
\end{equation}
The norm of the above quantity is the \textit{dissipation-to-driving ratio} $k$ [defined below Eq.~\eqref{eq:smax}]. It is a crucial quantity to determine the synchronization regime of the system (\textit{entrainment-dominant} or \textit{mutual-coupling dominant}) and in observing competition between these mechanisms in the engine regime [see Eq. \eqref{eq:smax}]. Solving for the populations gives us,
\begin{widetext}
\begin{eqnarray}
    \label{eq:rho_11_ss}
    \Rss_{11} &=& \frac{(1+n_h)\Big[2\lambda^2\big( n_c\gamma_c +\gamma_h(1+p)n_h\big)+\xi_h(1+p)\gamma_cn_c\big(\xi_c +\xi_h(1+p)\big)\Big]}{F(n_h, n_c, \gamma_h, \gamma_c, \lambda, p)}, \nonumber \\
    \label{eq:rho_22_ss}
    \Rss_{22} &=& \Rss_{33} = \frac{\lambda^2[n_h + n_c + 2n_h n_c+2\xi_h n_h(1+p)]+\xi_c\xi_h(1+p)n_h[\xi_c+\xi_h(1+p)]}{F(n_h, n_c, \gamma_h, \gamma_c, \lambda, p)},
\end{eqnarray}
\end{widetext}
with the denominator
\begin{widetext}
\begin{eqnarray}
    \label{eq:numerator_F}
    F(n_h, n_c, \gamma_h, \gamma_c, \lambda, p)&=&2\lambda^2\left[\gamma_c(1+3n_c+2n_h+4n_hn_c)+\xi_h(1+p)(1+4n_h)\right] \nonumber \\
    &&+\gamma_c\xi_h(1+p)(1+3n_h+2n_c+4n_h n_c)\left[\xi_c+\xi_h(1+p)\right],
\end{eqnarray}
\end{widetext}
On the other hand, using the population results, the steady state \textit{coherences} read,
\begin{equation}
    \label{eq: nondeg_coherences}
    \Rss_{12} = \Rss_{13} = i \frac{\lambda \gamma_c \xi_h (1+p)(n_c - n_h)}{F(n_h, n_c, \gamma_h, \gamma_c \lambda, p)},
\end{equation}
\begin{equation}
    \label{eq:deg_coherences}
    \Rss_{23} = \frac{\lambda^2 \gamma_c (n_c - n_h)}{F(n_h, n_c, \gamma_h, \gamma_c, \lambda, p)}.
\end{equation}
These steady state coherences are the important quantities in deriving the synchronization measure $S_{max}$ [see \eqref{eq:smax}]. The derivation of $S_{max}$ from the steady state coherences has been explained in detail in our accompanying work \cite{murtadho2023coexistence}. 
\section{Thermodynamic Observables in Steady State}\label{appendixB}
In Appendix \ref{appendixA}, we derived the expressions for the steady state of a degenerate four-level thermal maser with interference ($p \neq 0$). We can use those formulas to express thermodynamic observables such as power and heat current in the steady-state [see \eqref{eq:power}-\eqref{eq:cold_heat_current}], 
\begin{equation}
    \label{eq:pss_analytic}
    P^{ss} = -\frac{2\lambda^2\gamma_c\xi_h(\omega_2-\omega_1)(1+p)(n_h - n_c)}{F(n_h, n_c, \gamma_h, \gamma_c, \lambda, p)},
\end{equation}
\begin{equation}
    \label{eq:qcss_analytic}
    \dot{Q}_c^{ss} = -\frac{2\lambda^2\gamma_c\xi_h\omega_1(1+p)(n_h - n_c)}{F(n_h, n_c, \gamma_h, \gamma_c, \lambda, p)},
\end{equation}
\begin{equation}
    \label{eq:qhss_inc_analytic}
    \dot{Q}_h^{inc,ss} = \frac{2\lambda^2\omega_2\gamma_c\xi_h(n_h - n_c)}{F(n_h, n_c, \gamma_h, \gamma_c, \lambda, p)},
\end{equation}
\begin{equation}
    \label{eq:qhss_coh_analytic}
    \dot{Q}_h^{coh,ss} = \frac{2p\lambda^2\omega_2\gamma_c\xi_h(n_h - n_c)}{F(n_h, n_c, \gamma_h, \gamma_c, \lambda, p)}, 
\end{equation}
where the total heat current from the hot bath is the total of coherent and incoherent contribution $\dot{Q}_h^{ss} = \dot{Q}_h^{inc,ss} + \dot{Q}_h^{coh,ss}$. 
It is interesting to note that $\dot{Q}_h^{coh,ss}/\dot{Q}_h^{inc,ss} = p$. This relation gives a thermodynamic meaning to noise-induced coherence parameter $p$ as the ratio between coherent and incoherent heat current in the steady state. It also implies the suppression or enhancement of heat flow due to coherence is fully determined by the sign of $p$. When there is destructive interference ($p<0$), the direction of coherent heat flow is opposite to the incoherent one (suppression) whereas for constructive interference ($p>0$) they are in the same direction (enhancement).

Recall that the direction of power and heat flow determines whether the system operates as an engine or refrigerator. In the engine regime, heat flows from the hot bath to the cold bath, and power is produced ($\dot{Q}_h^{ss}>0, \; \dot{Q}_c^{ss}<0, \; P^{ss}<0$). Conversely, in the refrigerator regime, heat flows from the cold bath to the hot bath, and power is consumed ($\dot{Q}_h^{ss}<0, \; \dot{Q}_c^{ss}>0, \; P^{ss}>0$). Equations \eqref{eq:pss_analytic} - \eqref{eq:qhss_coh_analytic} implies that in the exact degenerate case ($\Delta = 0$), the engine-to-refrigerator transition is fully determined by the baths bosonic mean-occupation number. Specifically, $n_h>n_c$ ($n_h<n_c$) implies the maser is functioning as an engine (refrigerator). 

In the engine regime ($n_h>n_c$), the efficiency $\eta$ is calculated as
\begin{equation}
    \label{eq:eff_analytics}
    \eta = -\frac{P^{ss}}{\dot{Q}_h^{ss}} = 1-\frac{\omega_1
    }{\omega_2}\leq \eta_{C},
\end{equation}
where $\eta_C = 1-T_c/T_h$ is the Carnot's efficiency ($T_{c,h}$ are the temperatures of the cold and hot bath respectively). The Carnot bound can be obtained by writing $\omega_{1} = T_c \ln[(1+n_c)/n_c]$ and $\omega_{2} = T_h \ln[(1+n_h)/n_h]$. 

Meanwhile, in the refrigerator regime ($n_h<n_c$), we calculate the COP,
\begin{equation}
    \label{eq:cop_analytics}
    \chi = \frac{\dot{Q}_c^{ss}}{P^{ss}} = \frac{\omega_1}{\omega_2 - \omega_1}. 
\end{equation}
We emphasize that the simple forms of efficiency and COP derived in Eqs.~\eqref{eq:eff_analytics}-\eqref{eq:cop_analytics} are only valid in the degenerate limit ($\Delta =0$). In the main text, we lift the degeneracy and show that the efficiency and the COP are lower bounded by synchronization measure $S_{max}$. 
\end{document}